\documentclass[aps,pra,twocolumn,floats]{revtex4}

\usepackage{ifthen}
\usepackage{ifpdf}

\usepackage{latexsym}
\usepackage{amssymb}
\usepackage{bm}

\ifpdf
\usepackage{graphicx}
\usepackage{epstopdf}
\else
\usepackage{graphicx}
\usepackage{epsfig}
\fi

\newcommand{\eexp}{\mbox{e}^}

\newcommand{\tbox}[1]{\mbox{\tiny #1}}
 
\newcommand{\amatrix}[1]{\matrix{#1}} 

\newcommand{\be}[1]{\begin{eqnarray}\ifthenelse{#1=-1}{\nonumber}{\ifthenelse{#1=0}{}{\label{e#1}}}}
\newcommand{\ee}{\end{eqnarray}} 
\newcommand{\beq}{\begin{eqnarray}}
\newcommand{\eeq}{\end{eqnarray}}

\newcommand{\hide}[1]{}

\newcommand{\mpg}[2][1.0\hsize]{\begin{minipage}[b]{#1}{#2}\end{minipage}}

\begin{document}

\title
{Restricted quantum-classical correspondence \\
and counting statistics for a coherent transition}

\author{Maya Chuchem and Doron Cohen}

\affiliation{Department of Physics, Ben-Gurion University, Beer-Sheva 84105, Israel}

\begin{abstract} 
The conventional probabilistic point of view implies that if a particle 
has a probability $p$ to make a transition from one site to another site, 
then the average transport should be ${\langle Q \rangle=p}$  
with a variance ${\mbox{Var}(Q) =(1-p)p}$. In the quantum mechanical 
context this observation becomes a non-trivial manifestation   
of restricted quantum-classical correspondence. 
We demonstrate this observation by considering the full counting statistics 
which is associated with a two level coherent transition 
in the context of a continuous quantum measurement process.  
In particular we test the possibility of getting a valid result 
for ${\mbox{Var}(Q)}$ within the framework of the adiabatic picture, 
analyzing the simplest non-trivial example of a Landau-Zener crossing. 
\end{abstract}

\maketitle

\section{Introduction}

The discussion of quantum-classical correspondence (QCC) in the mathematical physics literature 
is as old as the history of quantum mechanics. Traditionally one has in mind complicated 
semiclassical approximation schemes for calculating the propagation of wavepackets. 
A more elaborated analysis of the spreading process~\cite{qcc}, leads to the distinction between 
robust {\em restricted QCC} and fragile {\em detailed QCC}.  
Restricted QCC means that ${\langle A\rangle_{\tbox{qm}} \approx \langle A\rangle_{\tbox{cl}}}$    
holds only for a restricted set of observables, and survives even in the presence 
of diffraction, while detailed QCC means ${\langle A\rangle_{\tbox{qm}} \approx \langle A\rangle_{\tbox{cl}}}$ 
for all the well behaved observables, and requires smooth potentials as in the traditional formulation.  \\

Controlling atoms in a few site system is state of the art \cite{raizen1,raizen2}. 
The prevailing studies are focused in restricted questions such 
as ``what is the probability for a transition from one site to the other site" \cite{niu1,niu2}. 
But what about the associated noise \cite{noise1,noise2} 
and the full counting statistics (FCS) of a quantum transition? 
We argue below that the simplest example 
of a single particle in a closed two site system can be worked out exactly   
so as to illuminate the essence of restricted QCC in the context 
of FCS studies. This has the side benefit of shedding new light on the vague picture 
in the pioneering works~\cite{levitov1,levitov2} 
about shot noise, where the discussion of FCS is immersed in complicated 
diagrammatic calculations involving a many-body system 
of Fermions in an open geometry, hence obscuring the simple physics 
that underlays the bottom line results.  
While most follow-up publications about FCS \cite{blanter,imry,lee,fano1,agam} 
are aimed in studying the distribution of transmission eigenvalues, 
our concern below is with some fundamental aspects  
that distinguish quantum dynamics from its classical stochastic analog.  \\

We consider the simplest non-trivial example: 
the adiabatic crossing of a particle from one site to another site, 
which involves a two level Landau-Zener transition \cite{zener,berry,fish,exact}. 
Within the framework of a classical probabilistic point of view 
the particle has some probability $p$ to make the transition. 
In any particular realization the particle either makes the transition or not. 
Accordingly the integrated current from the first to the second site 
is either ${Q=1}$ or ${Q=0}$ respectively. 
It follows that 
\be{1}
\langle \mathcal{Q} \rangle \ \ = \ \ p
\ee 
with the variance
\be{2}
\mbox{Var}(\mathcal{Q}) \ \ = \ \ (1-p)p
\ee  
This classical probabilistic point of view does not hold in the 
quantum mechanical reality. The characterization of the 
full counting statistics requires some care in the 
description of a continuous quantum measurement. 
Accordingly one has to distinguish between the naive 
mathematical definition \cite{levitov1} of the~$Q$ probability 
distribution $\mbox{P}(Q)$, and the proper 
physical definition \cite{nazarov} of the quasi-probability 
distribution $\mbox{P}(Q;x)$, where $x$ signifies 
the strength of the interaction with the detector. 
It is important to realize that in general $\mbox{P}(Q;x{=}0)$ 
is not the same as $\mbox{P}(Q)$, but still they have the 
same first and second moments. \\

This paper has two parts. In the first part 
we establish restricted QCC for the counting statistics 
using general principles. In the second part we 
study whether the adiabatic approximation can be used 
in order to derive leading order results. 
The importance of the latter analysis becomes apparent 
once one tries to obtain results for more complicated 
multiple-path geometries where exact solution 
for $\mbox{P}(Q)$ is out of reach. We further discuss 
this latter point in the concluding section.

\section{Restricted QCC for moments of $\mathcal{Q}$}

For a general network that consists of several sites 
we can define a set of occupation operators $\mathcal{N}$  
and current operators $\mathcal{I}$. 
In the Heisenberg picture the time derivative 
of any occupation operator equals 
a sum over the ingoing current operators.    
In order to demonstrate the general idea 
of restricted QCC we consider below the simplest 
example of one particle in a two-site system. 
The model Hamiltonian in the position basis is 
\be{3}
\mathcal{H} 
= \left( \begin{array}{cc}  \alpha t/2 & c \\  c & -\alpha t/2 \end{array} \right),
\ee 
where $c$ is the hopping amplitude. 
We define an occupation operator $\mathcal{N}$
and a current operator $\mathcal{I}$ as follows:
\be{44}
\mathcal{N} 
=  \left( \begin{array}{cc} 0 & 0 \\ 0 & 1 \end{array} \right).
\ee 
\be{4}
\mathcal{I} 
= \left( \begin{array}{cc} 0 & ic  \\ -ic & 0 \end{array} \right)
\ee
Using the Heisenberg picture the two are related by the equation 
\be{6}
\frac{d}{dt} \mathcal{N}(t) \ \ = \ \ \mathcal{I}(t) 
\ee 
where $\mathcal{I}(t)=U(t)^{\dag}\mathcal{I}U(t)$, 
and $U(t)$ is the evolution operator. 
Consequently it is useful to define 
a counting (``transported charge") operator:
\be{5}
\mathcal{Q} 
= \int_{0}^{t} \mathcal{I}(t') dt'  
\ee

From the Heisenberg equation of motion Eq.(\ref{e6}) 
it follows that the occupation probability $\langle \mathcal{N}(t) \rangle$
and the probability current $\langle \mathcal{I}(t) \rangle$
are related by a {\em continuity equation}. 
But in fact we can derive a stronger statement.
By integrating Eq.(\ref{e6}) over time 
one obtains 
\be{0}
\mathcal{Q} = \mathcal{N}(t)-\mathcal{N}(0) 
\ee 
Assuming that the particle is initially in the left site, 
which is a zero eigenstate of $\mathcal{N}$, 
one concludes the non-trivial relation
\be{51}
\langle \mathcal{Q}^k \rangle \ \ = \ \
\langle \mathcal{N}^k \rangle_t 
\ \ \ \ \ \ \ \ \ \ \mbox{for $k=1,2.$}
\ee
This relation between 
the $k$th moment of the counting operator 
(as determined in the Heisenberg picture by an expectation value at the reference time $t{=}0$) 
and the $k$th moment of the occupation operator 
(as determined in the Schrodinger picture by an expectation value at time~$t$) 
does not hold for ${k>2}$, because the operators in ${\langle 0|[\mathcal{N}(t)-\mathcal{N}(0)]^k| 0\rangle}$ 
are non-commuting.

Thus for a coherent quantum transition 
where initially  ${\langle \mathcal{N}^k \rangle_0=0}$, 
we find at the end of the process ${\langle \mathcal{N}^k \rangle_t=p}$,   
and consequently it follows that  ${\langle \mathcal{Q}^k \rangle=p}$ for ${k=1,2}$, 
leading to Eq.(\ref{e1}) and Eq.(\ref{e2}). But the higher moments  
would be different compared with the classical expectation, 
as further discussed in the next section, and therefore we have 
here a very simple example for {\em restricted} rather than detailed QCC.

\section{The FCS - naive results}

Both Eq.(\ref{e1}) and Eq.(\ref{e2}) can be derived 
on the basis of a classical probabilistic point of view. 
The classical reasoning is based on the following 
idea regarding the counting statistics:
\be{100}
\mbox{P}(Q) = 
\left\{\amatrix{
1{-}p  \ \ \ \ \ \ \ \ & \mbox{for} \ Q=0 \cr
p \ \ \ \ \ \ \ \ & \mbox{for} \ Q=1
}\right.
\ee
The $k$th moment of this probability distribution 
is ${\langle Q^k \rangle=p}$, leading 
to Eq.(\ref{e1}) as well as to Eq.(\ref{e2}). 
But we are going to explain that the classical reasoning 
is wrong: It is wrong both according to the 
naive mathematical definition of $\mbox{P}(Q)$, 
and also according to the proper physical 
definition of $\mbox{P}(Q;x)$, 
which we review in Appendix~A. 
Consequently in the present section 
we derive the mathematically correct 
results for $\mbox{P}(Q)$, while in the next 
section we work out the physically meaningful 
result for $\mbox{P}(Q;x)$.

Let us elaborate on the straightforward procedure for naive FCS calculation. 
The first step is to write the current operator  
in the Heisenberg picture:
\be{0}
\mathcal{I}(t)_{nm}
 \ \ = \ \ 
\langle n | U(t)^{\dag}\mathcal{I}U(t) | m \rangle 
\ee
It is important to realize that the current operator 
and hence the counting operator have a zero trace. 
Consequently integrating over time we get 
\be{0}
\mathcal{Q}_{nm} 
\ \ \equiv \ \ 
\left(\amatrix{
+Q_{\parallel} & iQ_{\perp} \cr
-iQ_{\perp}^* & -Q_{\parallel}
}\right)
\ee
The first two moments $\langle \mathcal{Q} \rangle$ and $\langle \mathcal{Q}^2 \rangle$
are obtained from this matrix, leading to the identifications:
\be{11}
\langle \mathcal{Q} \rangle \ \ &=& \ \ Q_{\parallel} 
\\ \label{e12}
\mbox{Var}(\mathcal{Q})  \ \ &=& \ \ |Q_{\perp}|^2 
\ee
The zero trace property also implies that the eigenvalues 
of $\mathcal{Q}$ are opposite in sign:
\be{130}
Q_{\pm} \ \ = \ \ \pm \sqrt{(Q_{\parallel})^2+|Q_{\perp}|^2}
\ee
On the basis of the continuity equation we can argue 
that ${\langle\mathcal{Q}\rangle_{\psi} \le 1}$ for any 
preparation. Therefore we must get ${|Q_{\pm}| \le 1}$. 
In fact we can deduce much more on the basis of restricted QCC. 
Observing that Eq.(\ref{e11}) and Eq.(\ref{e12}) 
should be in agreement with Eq.(\ref{e1}) and Eq.(\ref{e2}) 
one deduces that  
\be{14}
Q_{\parallel} \ \ &=& \ \ p
\\ 
\label{e260}
Q_{\perp} \ \ &=& \ \ \sqrt{(1{-}p)p} \ \times \ \mbox{PhaseFactor}
\ee

If we could regard $\mathcal{Q}$ 
as a conventional observable,  
then upon measurement its observed values 
would have the distribution 
\be{101}
\mbox{P}(Q) = 
\left\{\amatrix{
p_{-} \ \ \ \ \ \ \ \ & \mbox{for} \ Q=Q_{-} \cr
p_{+} \ \ \ \ \ \ \ \ & \mbox{for} \ Q=Q_{+}
}\right.
\ee
This distribution is characterized by two parameters 
because $Q_{\pm}$ are opposite and sign, 
and $p_{\pm}$ sum up to unity. Restricted QCC 
provides the two equations  ${\langle \mathcal{Q} \rangle=p}$ 
and ${\langle \mathcal{Q}^2 \rangle=p}$ 
that can be solved, leading to 
\be{102}
Q_{\pm} &=& \pm\sqrt{p} \\
p_{\pm} &=& \frac{1}{2}\left(1\pm\sqrt{p}\right)
\ee
where the expression for $Q_{\pm}$ is in agreement 
with Eqs.(\ref{e130})-(\ref{e260}) of the previous paragraph. 
A look-alike result had been obtained for shot noise 
of Fermions using complicated diagrammatic techniques~\cite{levitov1}.  
An equivalent way to express Eq.(\ref{e101}) is to say that the $k$th moment is 
\be{103}
\langle \mathcal{Q}^k \rangle = p_{+}Q_{+}^k +p_{-}Q_{-}^k = p^{\left\lfloor\frac{k+1}{2}\right\rfloor}
\ee 
where $k=0,1,2,3,...$ and $\lfloor...\rfloor$ stands 
for the integer part (i.e. rounded downwards).
The corresponding classical result is~$p$ irrespective of~$k$.
One may say that we have encountered here the simplest 
example of {\em restricted} QCC.

\section{The FCS - physical results}

In a later publication \cite{levitov2} the naive mathematical 
definition of the full counting statistics has been criticized. 
The most illuminating approach \cite{nazarov} is to analyze the reduced dynamics 
of the detector. Using the standard von-Neumann pointer scheme \cite{pointer}
and transforming to the Wigner representation one deduces  
that the final state $\rho(q,x)$ of the pointer is the convolution 
of its initial state with a kernel  $\mbox{P}(q{-}q';x)$, 
as defined in Eq.(\ref{e1000}).  It follows that 
\be{13}
&& \mbox{P}(Q;x{=}0) \ = 
\\ \nonumber
&& \frac{1}{2\pi} \int 
\left\langle 
\left[ \mathcal{T}\eexp{-i(r/2)\mathcal{Q}} \right]^{\dag} 
\left[ \mathcal{T}\eexp{+i(r/2)\mathcal{Q}} \right] 
\right\rangle 
\eexp{-iQr} dr
\ee
can be regarded as a quasi-distribution that describes 
the full quantum statistics. It is of course a physically 
measurable object. The naive mathematical definition  
is obtained if we ignore time ordering:
\be{-1}
\mbox{P}(Q) 
\ \ &=& \ \ 
\frac{1}{2\pi} \int 
\left\langle 
\eexp{+ir\mathcal{Q}} 
\right\rangle 
\eexp{-iQr} dr
\\
\ \ &=& \ \ 
\left\langle \delta(Q-\mathcal{Q})\right\rangle
\ee

In general the calculation of $\mbox{P}(Q;x{=}0)$ 
is very complicated. But we can gain some insight 
by considering the simplest case of a Bloch transition, 
which is Eq.(\ref{e3}) with $\alpha=0$. 
After time~$t$ the probability to find the 
particle in the second site is ${p=[\sin(ct)]^2}$. 
The expectation value in the integrand of Eq.(\ref{e13}) is 
\be{0}
\tilde{\mbox{P}}(r)=1-\frac{r}{(r/2)+i} \left[\sin\left(ct\sqrt{1+(r/2)^2}\right)\right]^2
\ee
[More conveniently one can use instead of  Eq.(\ref{e13}) 
the equivalent expression Eq.(\ref{e1000}), 
where the $U$s can be interpreted as spin~1/2 rotation matrices].
In order to get $\mbox{P}(Q;x{=}0)$ we have to Fourier transform (FT) 
the function $\tilde{\mbox{P}}(r)$. For the purpose of discussion let us assume 
long times (${ct\gg1}$) so as to have separation of scales.   
Then we can distinguish between a central part 
where ${\tilde{\mbox{P}}(r)\approx 1+ipr-(1/2)pr^2+...}$  as implies by restricted QCC, 
and oscillatory far tails where ${\tilde{\mbox{P}}(r)\approx \cos(ctr)}$. 
The FT of the central part gives a non-singular exponential-like piece, 
while the FT of the tails contributes two delta functions ${(1/2)\delta(Q \pm ct)}$ 
which are screened by negative clouds as illustrated in Fig.~1. 
We can regard the singular delta functions at $Q=\pm ct$ 
as the remnants of those that are centered at $Q_{\pm}=\pm\sqrt{p}$ 
in the naive calculation. It is interesting that for long time 
their location reflects the eigenvalues of the current operator 
rather than the eigenvalues of the counting operator.

Fig.~2 is a caricature that illustrates the significance 
of the FCS results with regard to the outcome of a quantum measurement. 
We sketch the final probability distribution of the Von Neumann pointer 
as implied by different ``conceptions"  .
The upper panel is based on the classical expectation 
of having either ${Q=0}$ or ${Q=1}$ displacement. 
The middle panel is the naive quantum mechanical prediction 
that suggests ${Q=Q_{\pm}}$ displacements, 
while the lower panel is the actual quantum prediction 
based on a convolution with $\mbox{P}(Q;x)$.  
An actual numerical illustration for the outcome of this 
convolution is presented in Fig.~3. Needless to say that 
in principle one can measure not only the $q$~distribution 
but the whole probability matrix so as to determine $\mbox{P}(Q;x)$ 
and in particular the quasi probability distribution $\mbox{P}(Q;x{=}0)$ 
via a deconvolution procedure. But looking in Fig.~3, 
it is nice to realize that the main features of the FCS 
are not smeared and pop to the eyes even without any deconvolution.

\section{FCS in the adiabatic approximation}

In the following sections we would like to examine the 
capabilities of the {\em leading order} adiabatic approximation 
in obtaining physically significant results for $\mbox{Var}(\mathcal{Q})$.
We consider the model system of Eq.(\ref{e3}). 
Initially (at $t{=}-\infty$) the particle is 
in the ground state, which is the left site. 
The probability at $t{=}\infty$ to find 
the particle in the right site is 
\be{0}
p \ \ = \ \ 1-P_{\tbox{LZ}} 
\ee
where
\be{230}
P_{\tbox{LZ}} \ \ = \ \ 
\exp{\left[ -2\pi\frac{c^2}{\alpha}  \right]} 
\ \ \equiv \ \ 
\exp{\left[ -\frac{\pi}{2}\gamma  \right]}
\ee
is the Landau-Zener probability to make 
a transition from the lower to the upper energy level. 
The rate of the driving is characterized 
by the dimensionless parameter 
\be{0}
\gamma \ \ = \ \ (2c)^2/\alpha.  
\ee
The straightforward way to calculate $\mbox{Var}(Q)$ is also 
the most complicated one, because it requires an explicit 
evaluation of the evolution operator~$U(t)$. 
For our two site system this can be done using parabolic cylinder functions~\cite{exact}. 
But having in mind more complex systems \cite{cnb}, 
for which exact solution are not available,   
we would like to make the explicit calculation 
within the framework of the adiabatic approximation:  
\be{-1}
U(t) \ \approx \ 
\sum_n \Big|n(t)\Big\rangle 
\ \exp\left[-i\int_{t_0}^{t} E_n(t')dt'\right] \ 
\Big\langle n(t_0)\Big|
\ee
In this expression $|n(t)\rangle$ and $E_n(t)$ are 
the so-called adiabatic states and the adiabatic energies \cite{berry}.  
The associated expression for the matrix elements of 
the time dependent current operator in the Heisenberg picture is: 
\be{-1}
\mathcal{I}(t)_{nm}
 \ \ &=& \ \ 
\langle n | U(t)^{\dag}\mathcal{I}U(t) | m \rangle 
\\ \nonumber
\ \ &\approx& \ \ 
\langle n(t) |\mathcal{I}| m(t) \rangle 
\times \exp\left[i\int_{t_0}^{t} E_{nm}(t')dt'\right]
\\
\ \ &=& \ \ 
\left(\amatrix{... & ic\eexp{i\Phi(t)} \cr  -ic\eexp{-i\Phi(t)} & ...} \right)
\label{e250}
\ee
where
\be{0}
\Phi(t) 
\ \ = \ \ \int^t \sqrt{(\alpha t)^2 + (2c)^2} \, dt'
\ee
Using the zero order ($\mathcal{O}(\alpha^0)$) adiabatic states 
we get for the off diagonal terms ${Q_{\perp} \approx Q_{\tbox{LZ}}}$ where 
\be{-1}
Q_{\tbox{LZ}} \ \ &\equiv& \ \ 
\frac{\gamma}{2}\int_{-\infty}^{\infty}\,\eexp{i\Phi(\tau)} d\tau
\\ \label{e28} 
\ \ &=& \ \
\frac{\gamma}{2}\int_{-\infty}^{\infty} \cosh(z) \,\eexp{i\Phi(z)} dz
\ee
Above we use the substitution ${t=(2c/\alpha)\tau}$ 
followed by ${\tau=\sinh(z)}$.

Before we go on with the calculation of $Q_{\tbox{LZ}}$ 
we would like to comment on the calculation of 
the diagonal terms in Eq.(\ref{e250}). This terms are 
trivially related via Eq.(\ref{e14}) to the 
Landau-Zener transition probability $P_{\tbox{LZ}}$ of Eq.(\ref{e230}).  
The leading order estimate of this probability 
is based on a conventional time dependent treatment \cite{fish}, leading to 
\be{-1}
P_{\tbox{LZ}}
\ \ &\approx& \ \ 
\left|\frac{1}{2}\int_{-\infty}^{\infty}\frac{1}{\tau^2+1}\,\eexp{i\Phi(\tau)} d\tau\right|^2
\\ \label{e29} 
\ \ &=& \ \ 
\left|\frac{1}{2}\int_{-\infty}^{\infty} \frac{1}{\cosh(z)} \,\eexp{i\Phi(z)} dz\right|^2
\ee
It is important to realize that in the zero 
order adiabatic approximation we get {\em zero}
for the diagonal terms of  Eq.(\ref{e250}), 
because the zero order adiabatic states are time-reversal symmetric 
and hence do not support non-zero average current.
If we use the first order ($\mathcal{O}(\alpha^1)$) adiabatic states 
we get for the diagonal terms $\pm1$ but still miss 
the non-adiabatic $P_{\tbox{LZ}}$ correction.
It is therefore non-trivial and requires verification  
that ${Q_{\perp} \approx Q_{\tbox{LZ}}}$ is a valid approximation.

\section{The calculation of $Q_{\tbox{LZ}}$}

In order to evaluate $Q_{\tbox{LZ}}$ of Eq.(\ref{e28})
we use the contour integration method as in Ref.\cite{fish}.
The explicit expression for the phase $\Phi$ as a function of ${z=x+iy}$ is 
\be{0}
\Phi(t) 
\ \ &=&  \ \ 
\frac{\gamma}{2} \left(z+\frac{1}{2}\sinh(2 z)\right) 
\\ \nonumber
\nonumber
\ \ &=& \ \ 
\frac{\gamma}{2}
\left[\left(x+\frac{1}{2}\sinh(2 x)\cos(2y)\right)\right.
\\ \nonumber
&&\left. +i\left(y+\frac{1}{2}\cosh(2 x)\sin(2y)\right)\right]
\ee
The contour of integration in Eq.(\ref{e28}) is $y=0$, 
but we would like to deform it into the complex plane 
so as to get rid of the rapid oscillations of the phase factor, 
and have instead a smooth monotonic variation. 
The deformed contour is displayed in Fig.~4.  
The phase is pure imaginary along the 
curves $C_{-}$ and $C_{+}$. At ${z_0=0+i(\pi/2)}$ 
we have ${\Phi=i(\pi/4)\gamma}$, 
while $\cosh{z} \approx i(z-z_0)$. 
Consequently in the $P_{\tbox{LZ}}$ integral of Eq.(\ref{e29})
we have a pole leading to the standard LZ results 
(disregarding the prefactor which requires higher orders).  
But in the case of the $Q_{\tbox{LZ}}$ integral Eq.(\ref{e28})
we do not have a pole: rather we have to consider the 
non singular part that comes from the integration 
along the $C_{\pm}$ curves. One observes that 
\be{-1}
Q_{\tbox{LZ}} 
\ \ &=& \ \ \int_{C_{-}+C_{+}}\!\!\!\!\!\!\!\!\!\!\!\!\!\!\!... dz 
\ \ = \ \ 2\Re \int_{C_{+}}\!\!\!\!\!\!... dz 
\\ \nonumber
\ \ &=& \ \ \int_0^{\infty} f(x) \ \eexp{i\Phi(x)} dx
\ee
where 
\be{-1} 
i\Phi(x) &=& -\frac{\gamma}{4}\left[\pi-\arccos\left(\frac{2x}{\sinh(2x)}\right) \right. 
\\ \nonumber
&&\left. +\left[1-\left(\frac{2x}{\sinh(2x)}\right)^2\right]^{1/2}\cosh(2x)\right] 
\ee
and
\be{-1}
f(x) = \frac{\gamma}{\sqrt{2}}\left[1{-}\frac{2x}{\sinh(2x)}\right]^{-1/2} \left(\frac{\sinh(4x)-4x}{\cosh(4x)-1}\right) \sinh(x)
\ee
Deep in the adiabatic regime (${\gamma\gg1}$) the integration is 
dominated by the small~$x$ interval where ${f(x)\propto \gamma x}$
and ${[\Phi(x){-}\Phi(0)]\propto \gamma x^3}$. Accordingly 
\be{0}
Q_{\tbox{LZ}} \ \ \sim \ \ \gamma^{1/3}\,\exp{\left[ -\frac{\pi}{4}\gamma  \right]}
\ee
Inspired by the calculation procedure of ${P_{\tbox{LZ}}}$ in Ref.\cite{berry}, 
our speculation is that also here the pre-exponential term 
would be renormalized to unity by higher orders, 
which would imply consistency with the expected (exact) 
result ${|Q_{\perp}|= \sqrt{(1-P_{\tbox{LZ}})P_{\tbox{LZ}}}}$ 
that follows from Eq.(\ref{e260}).

\section{Discussion}

We have demonstrated how restricted QCC can be established 
and utilized in order to determine the counting statistics 
of a quantum coherent transition. It is important to realize 
that this procedure has some limitations. Namely,  
if we had considered not a two site system but a more complex system 
with multiple path geometry, then the same considerations 
would not allow to deduce the variance $\mbox{Var}(\mathcal{Q})$ 
of the integrated current. In particular one wonders 
what happens if a particle has {\em two} optional paths 
available to get from one site to another site \cite{heiblum}. 
Within the framework of the classical probabilistic 
theory, a splitter would imply a noisy outgoing current.  
But in Ref.\cite{cnb} we argue that the coherent 
splitting of a wavepaket is not noisy, and furthermore  
that the whole study of fluctuations in quantum stirring 
devices requires to go beyond QCC considerations.  
Consequently the problem that has been raised in Ref.\cite{cnb} 
has motivated us to study the capabilities of the leading order 
adiabatic approximation for the purpose of calculating $\mbox{Var}(\mathcal{Q})$.

The calculation of the Landau-Zener transition 
probability ${P_{\tbox{LZ}}}$ 
and the associated dispersion ${Q_{\tbox{LZ}}}$  
in the leading order adiabatic approximation 
yields a contour integral in the complex plane.
While the ${P_{\tbox{LZ}}}$ integral 
is dominated by a pole, the ${Q_{\tbox{LZ}}}$
integration is related to the corresponding 
principal part. Restricted QCC implies 
that ${P_{\tbox{LZ}}}$ and ${Q_{\tbox{LZ}}}$ 
are related. Our analysis has demonstrated that 
the dominant exponential term is correctly reproduced, 
but not the pre-exponential term which apparently    
requires infinite order.

The traditional QCC principle is based 
on the semiclassical approximation and implies 
{\em detailed} QCC for all the moments up to 
the quantum resolution limit.   
In contrast to that {\em restricted} QCC 
is very robust, and survives even 
in the presence of diffraction \cite{qcc}. 
The Landau-Zener problem is possibly the simplest 
non-trivial example where this idea can be demonstrated. 
For some more complicated (chaotic) systems 
restricted QCC can be established as an approximation 
in the long time limit, on the basis 
of the short time perturbative analysis  
which is extrapolated using the central limit theorem.   
It would be interesting to explore 
the implications of coupling to the environment 
in this context \cite{efrat}.



\ \\ \ \\ 

{\bf Acknowledgments --}
The original motivation to deal with counting statistics 
in the context of closed systems came from a discussion 
with Oded Agam (HUJI). Yuly Nazarov (Delft) is acknowledged 
for highlighting some major observations regarding 
full counting statistics following Refs.\cite{levitov1,levitov2,nazarov}. 
DC thanks Miriam Blaauboer for a most fruitful visit in Delft TU, 
during which this work has been crystallized. 
We also enjoyed talking with Yaroslav Blanter (Delft), 
Yuval Gefen (Weizmann), and Shmuel Fishman (Technion). 
The research was supported by grants from  
the Deutsch-Israelische Projektkooperation (DIP), 
and the US-Israel Binational Science Foundation (BSF).

\appendix

\section{The Von-Neumann measurement scheme}

In this section we present a short simple derivation 
of the main result of Ref.\cite{nazarov} regarding 
the measurement of the full quantum statistics. 
The original derivation has been based on 
an over-complicated path integral approach.

The coupling of the system to a Von-Nuemann 
pointer \cite{pointer} whose canonical coordinates are ${(\hat{x},\hat{q})}$ 
is described by the Hamiltonian 
\be{0}
\mathcal{H}_{\tbox{total}} = \mathcal{H}(t) - \mathcal{I}\hat{x}
\ee
The states of system can be expanded in some basis $|n\rangle$,  
and accordingly for the system with the detector 
we can use the basis ${|n,x\rangle}$. 
The representation of the evolution operator is  
\be{0}
U(n,x|n_0,x_0) \ \ = \ \ U[x]_{n,n_0} \ \delta(x-x_0)
\ee
where $U[x]$ is a system operator that depends on the constant parameter~$x$.    
We formally write its explicit expression both in the Schrodinger picture
and also in the interaction picture using time ordered exponentiation:  
\be{-1}
U[x] 
\ \ &=& \ \ \mathcal{T}\exp\left[-i\int_0^t (\mathcal{H} - x\mathcal{I})dt' \right]   
\\
\ \ &=& \ \ U[0] \,\mathcal{T}\exp\left[i x \int_0^t \mathcal{I}(t')dt' \right]  
\ee
The time evolution of the detector is described by its 
reduced probability matrix 
\be{-1}
&& \rho_t(x'',x') \ \ = \ \ 
\sum_{n, n_0',n_0''}\int dx_0'dx_0''  
\\ \nonumber
&& U(n,x''|n_0'',x_0'')  U(n,x'|n_0',x_0')^* 
\, \rho^{\tbox{sys}}_{n_0'',n_0'} \rho_0(x_0'',x_0')   
\\ \nonumber
&& = \left[\sum_{n,n_0',n_0''} 
U[x'']_{n,n_0''} U[x']_{n,n_0'}^* 
\, \rho^{\tbox{sys}}_{n_0'',n_0'} \right] \,\, \rho_0(x'',x')   
\ee
where ${\rho^{\tbox{sys}} = |\psi\rangle\langle\psi|}$ is the 
initial state of the system, and ${\rho_0(x_0'',x_0')}$ 
is the initial preparation of the detector. Transforming to 
the Wigner function representation we get the convolution  
\be{999}
\rho_t(q,x) = \int \mbox{P}(q-q';x) \rho_0(q',x) dq'
\ee
where 
\be{1000}
&& \mbox{P}(Q;x) \ \ = \ \ 
\\ \nonumber
&& \frac{1}{2\pi} \int 
\left\langle \psi \Big| U[x-(r/2)]^{\dag} U[x+(r/2)] \Big| \psi \right\rangle \eexp{-iQr} dr
\ee

\newpage


\clearpage

\mpg{
\begin{center}
\includegraphics[width=8cm]{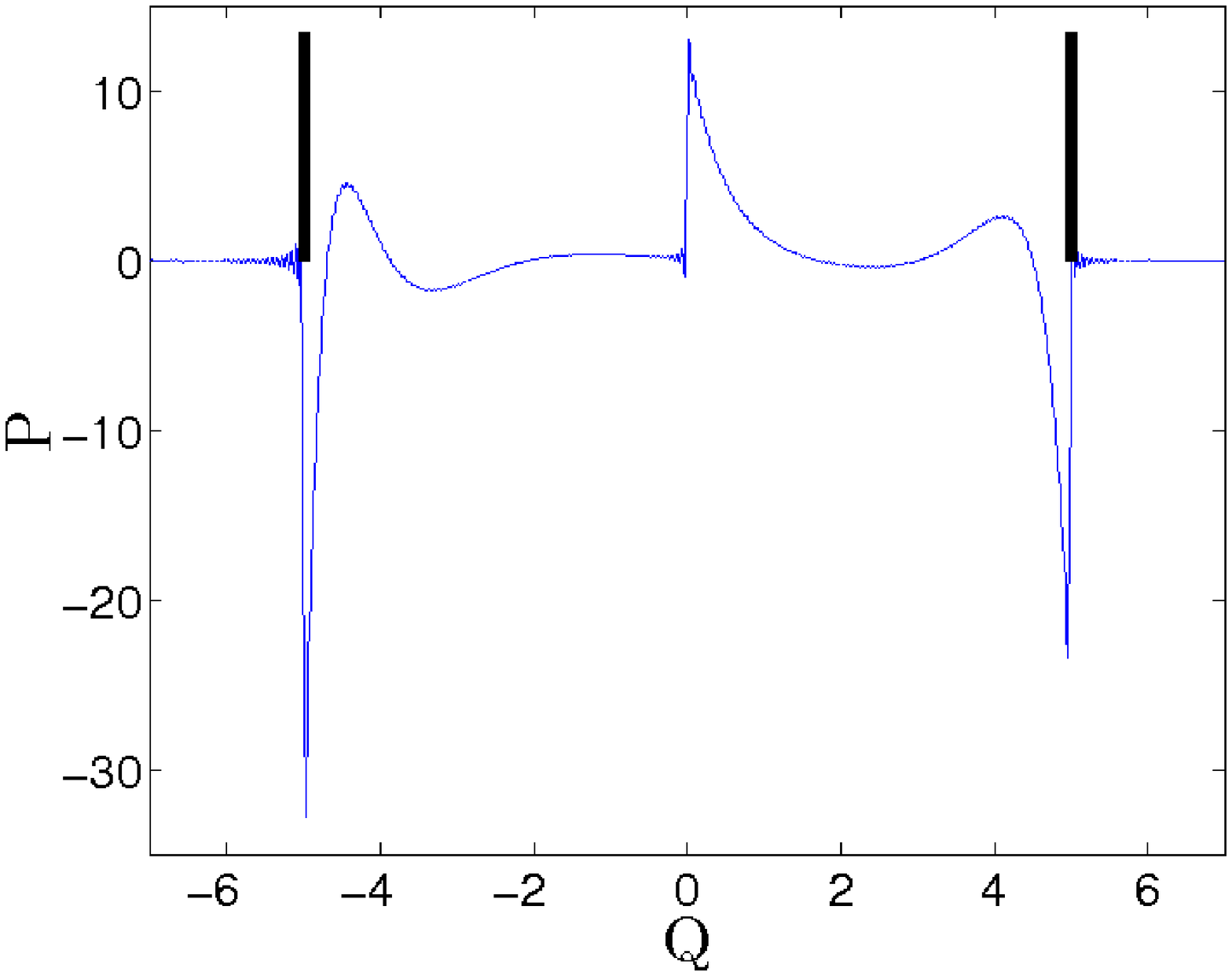}
\end{center}

{\footnotesize 
{\bf Fig.1:} Numerical calculation of the full counting 
statistics kernel $\mbox{P}(Q;x{=}0)$ for a coherent 
Bloch transition from one site to another site. 
The evolution is governed by the Hamiltonian of Eq.(\ref{e3}) 
with ${\alpha=0}$ and ${ct=5}$. The thick bars represent 
delta functions. See the text for further explanations.
}
}

\ \\ \ \\

\mpg{
\begin{center}
\includegraphics[width=8cm]{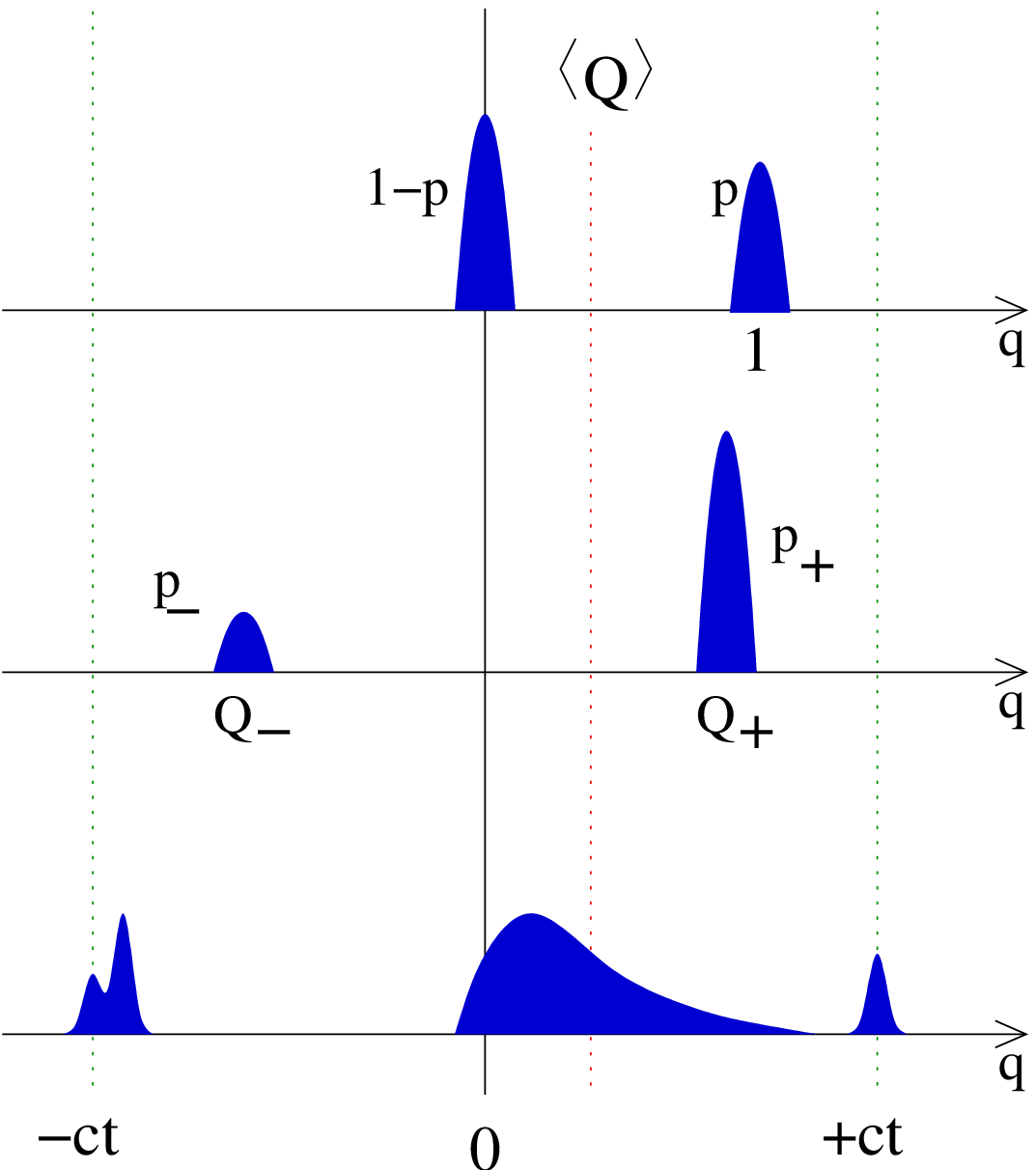}
\end{center}

{\footnotesize 
{\bf Fig.2:} A caricature that illustrates 
the final probability distribution of the Von Neumann 
pointer~$q$. It is assumed that the initial state 
is a wavepacked that is concentrated at ${q \sim 0}$.    
The upper panel is based on the classical expectation 
of having either ${Q=0}$ or ${Q=1}$ displacement. 
The middle panel is the naive quantum mechanical prediction 
that suggests ${Q=Q_{\pm}}$ displacements, 
while the lower panel is the actual quantum prediction 
based on a convolution with $\mbox{P}(Q;x)$.  
}
}

\newpage

\mpg{
\begin{center}
\includegraphics[width=8cm]{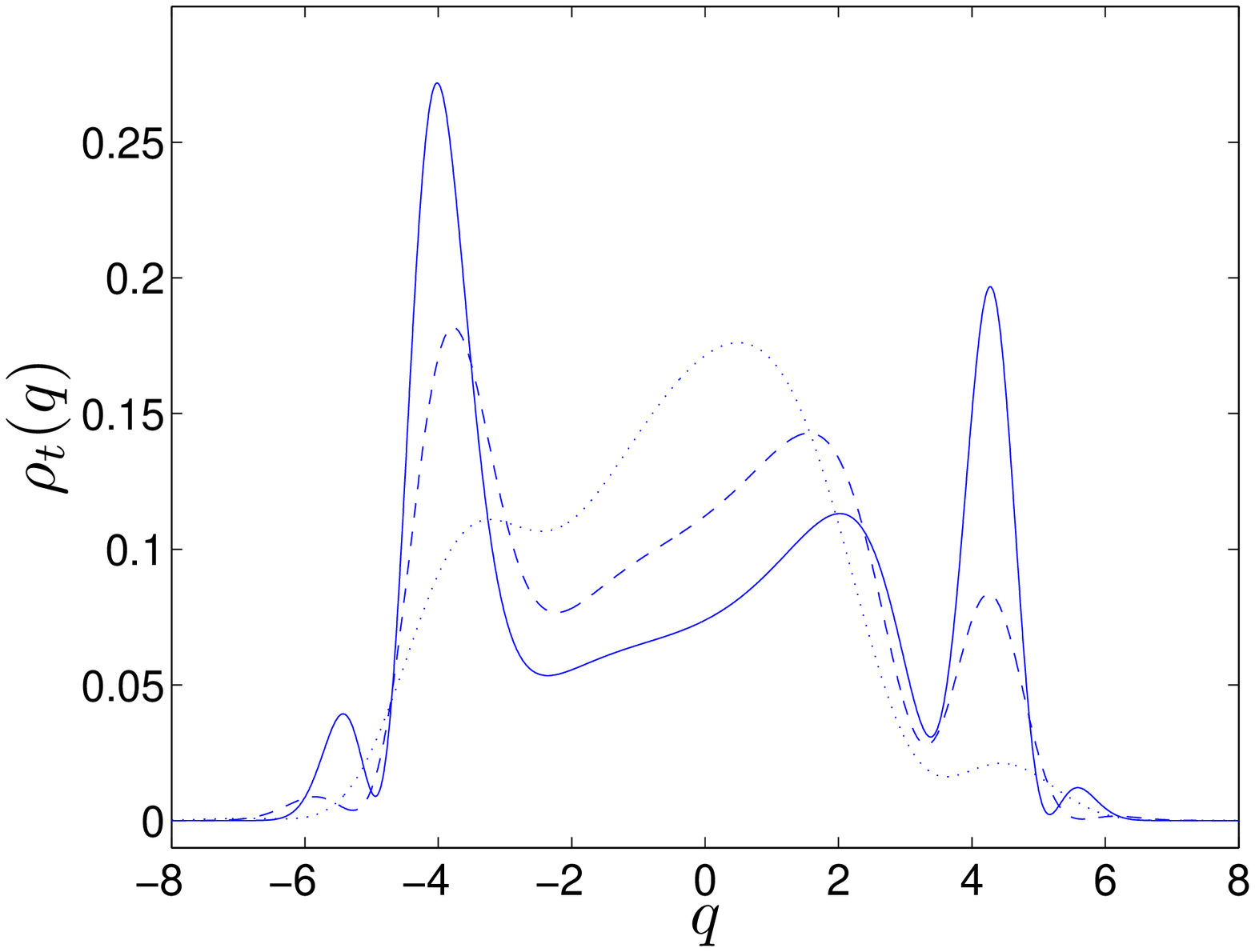}
\end{center}

{\footnotesize 
{\bf Fig.3:} Numerical calculation that illustrates 
the final probability distribution of the Von Neumann 
pointer~$q$ corresponding to the lower panel in Fig.~2.
The parameters are the same as in Fig.~1. 
The initial preparation is a Gaussian wavepacket 
centered at ${q=x=0}$ of 
width ${\sigma_x=1.2}$ (solid line), 
and ${\sigma_x=0.8}$ (dashed line), 
and ${\sigma_x=0.5}$ (dotted line). 
Its evolution has been calculated using Eq.(\ref{e999}) 
with Eq.(\ref{e1000}).}
}

\ \\

\mpg{
\begin{center}
\includegraphics[width=8cm]{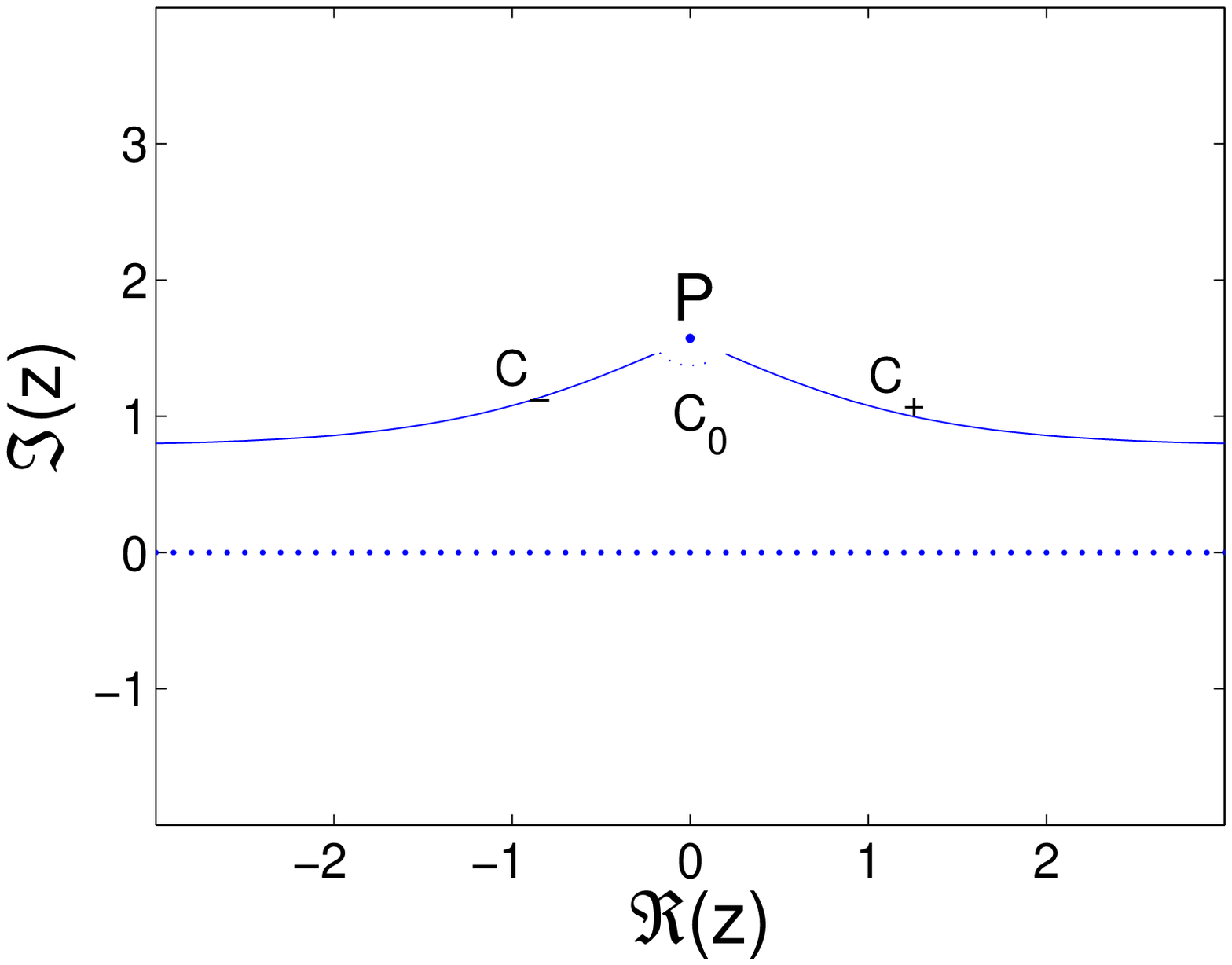}
\end{center}

{\footnotesize 
{\bf Fig.4:} 
The calculation of ${P_{\tbox{LZ}}}$ and  ${Q_{\tbox{LZ}}}$
is done by deforming the integration contour into the 
complex plane. The integration is along the 
curves $C_{\pm}$ where the integrand is non-oscillatory, 
and along the arc $C_{0}$ that encircles the pole~$P$ 
at ${z_0=0+i(\pi/2)}$. The contributions of 
the (non-displayed) vertical segments that 
connect $C_{-}$ and $C_{+}$ 
to the real axis at infinity can be neglected.
}
}

\end{document}